\input{aipcheck}

\documentclass[
    ,final            
  ]
  {aipproc}

\layoutstyle{6x9}

\begin{document}

\title{HEAVY FLAVOR DECAYS AND LIGHT HADRONS IN THE FOCUS EXPERIMENT: RECENT RESULTS}


\author{Sandra Malvezzi\thanks{on behalf of the FOCUS Collaboration
(http://www-focus.fnal.gov/)} }{
  address={INFN Sezione di Milano Bicocca,P.za della Scienza 3, 20126 Milano Italy}
}

\begin{abstract}
Interpretation of $D$-meson decay-dynamics has revealed itself to be strongly
dependent on our understanding of the light-meson sector. The interplay
becomes particularly evident in Dalitz plot analyses to study physics within
and beyond the Standard Model. Experience and results from FOCUS are presented
and discussed. A brief update of the pentaquark search in the experiment is
also reported.

\end{abstract}

\maketitle


\section{Introduction}

Dalitz analyses are largely applied in modern high-energy experiments to study
Heavy Flavor hadronic decays, but also to perform precise measurements of the
CKM matrix elements and search for new physics. Paradigmatic examples are $B
\to \rho \pi$ and $B \to D^{(*)}K^{(*)}$ for the extraction of the $\alpha$
and $\gamma$ angles of the Unitarity Triangle. The extraction of $\alpha$ in
$B \to \rho \pi$ requires filtering the desired intermediate states among all
the possible $(\pi\pi)\pi$ combinations, e.g. $\sigma \pi$, $f_0(980) \pi$
etc. The extraction of $\gamma$ in $B \to D^{(*)}K^{(*)}$ requires, in turn,
modeling the $D$ amplitudes. The $\pi \pi$ and $K\pi$ S-wave are characterized
by broad, overlapping states: unitarity is not explicitly guaranteed by a
simple sum of Breit--Wigner functions. In addition, independently of the
nature of the $\sigma$, it is not a simple Breit--Wigner. The $f_0(980)$ is a
Flatt\'e-like function, and its lineshape parametrization needs precise
determination of $K K$  and $\pi\pi$ couplings. Recent analyses of CP
violation in the $B \to D K$ channel from the beauty factories \emph{needed}
two \emph{ad hoc} resonances to reproduce the excess of events in the $\pi\pi$
spectrum, one at the low-mass threshold, the other at 1.1\,GeV$^2$
\cite{gamma_babar,gamma_belle}. This procedure of ``effectively'' fitting data
invites a word of caution on estimating the systematics of these measurements.
A question then naturally arises: in the era of precise measurements, do we
know sufficiently well how to deal with strong-dynamics effects in the
analyses? We have faced parametrization problems in FOCUS and learnt that many
difficulties are already known and studied in different fields, such as
nuclear and intermediate-energy physics, where broad, multi-channel,
overlapping resonances are treated in the \emph{K-matrix} formalism
\cite{wigner,chung, aitch}. The effort we have made consisted mainly in
building a bridge of knowledge and language to reach the high-energy
community; our pioneering work in the charm sector might inspire future
accurate studies in the beauty sector. FOCUS Dalitz plot analyses of the
$D+,D_s \to \pi^+ \pi^- \pi^+$ and of the $D^+ \to K^-\pi^+\pi^+$ will be
discussed.

The collaboration has also taken a complementary non-parametric approach to
measuring the $K^-\pi^+$ amplitude in the $D^+\to K^-K^+\pi^+$ decay using a
projective weighting technique. Results will be presented.

\section{The $D^+$ and $D_s \to \pi^+ \pi^- \pi^+$ amplitude analysis} 

\begin{figure}[h]
\centering 
\includegraphics[width=70mm]{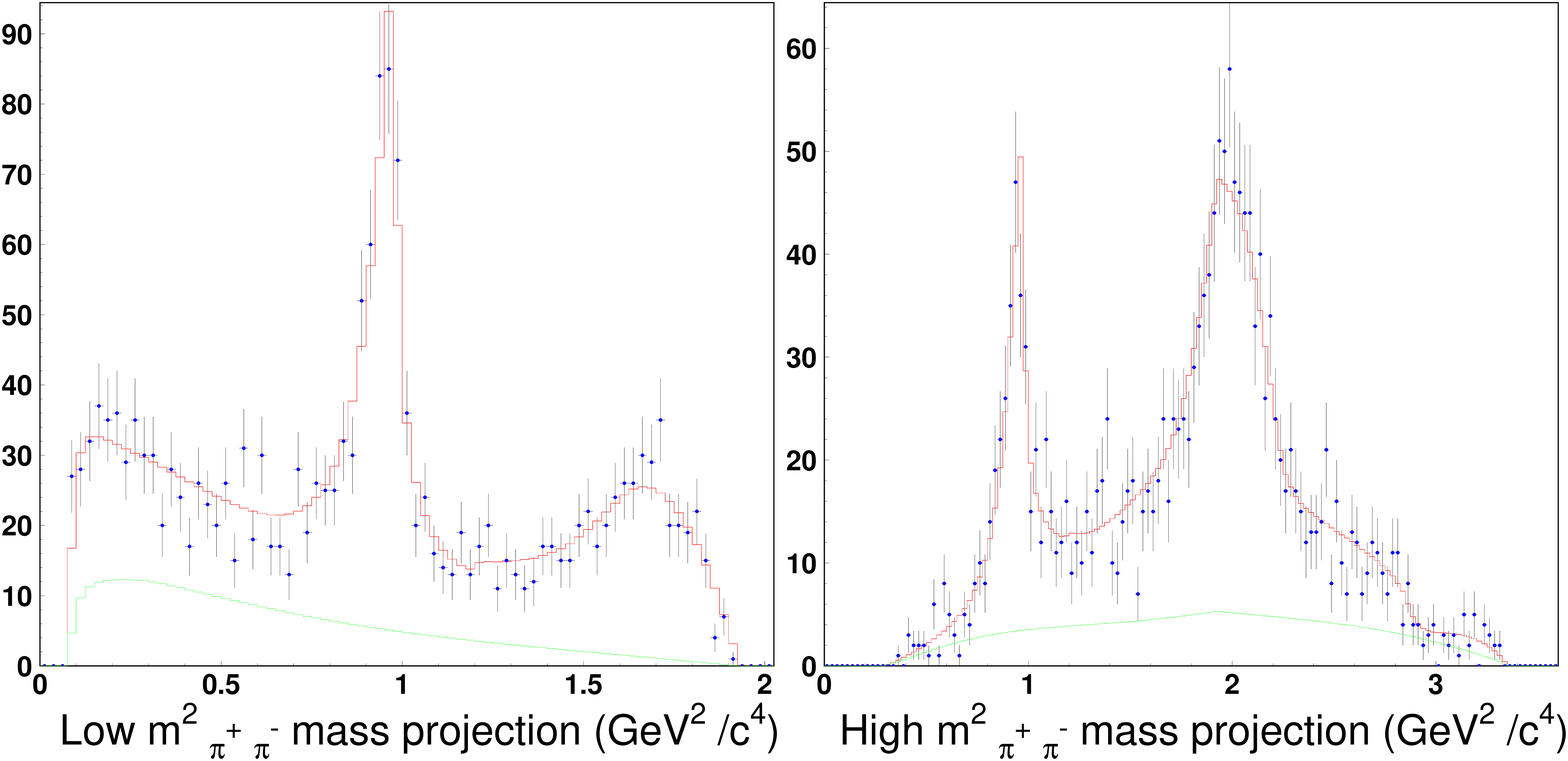}
\includegraphics[width=70mm]{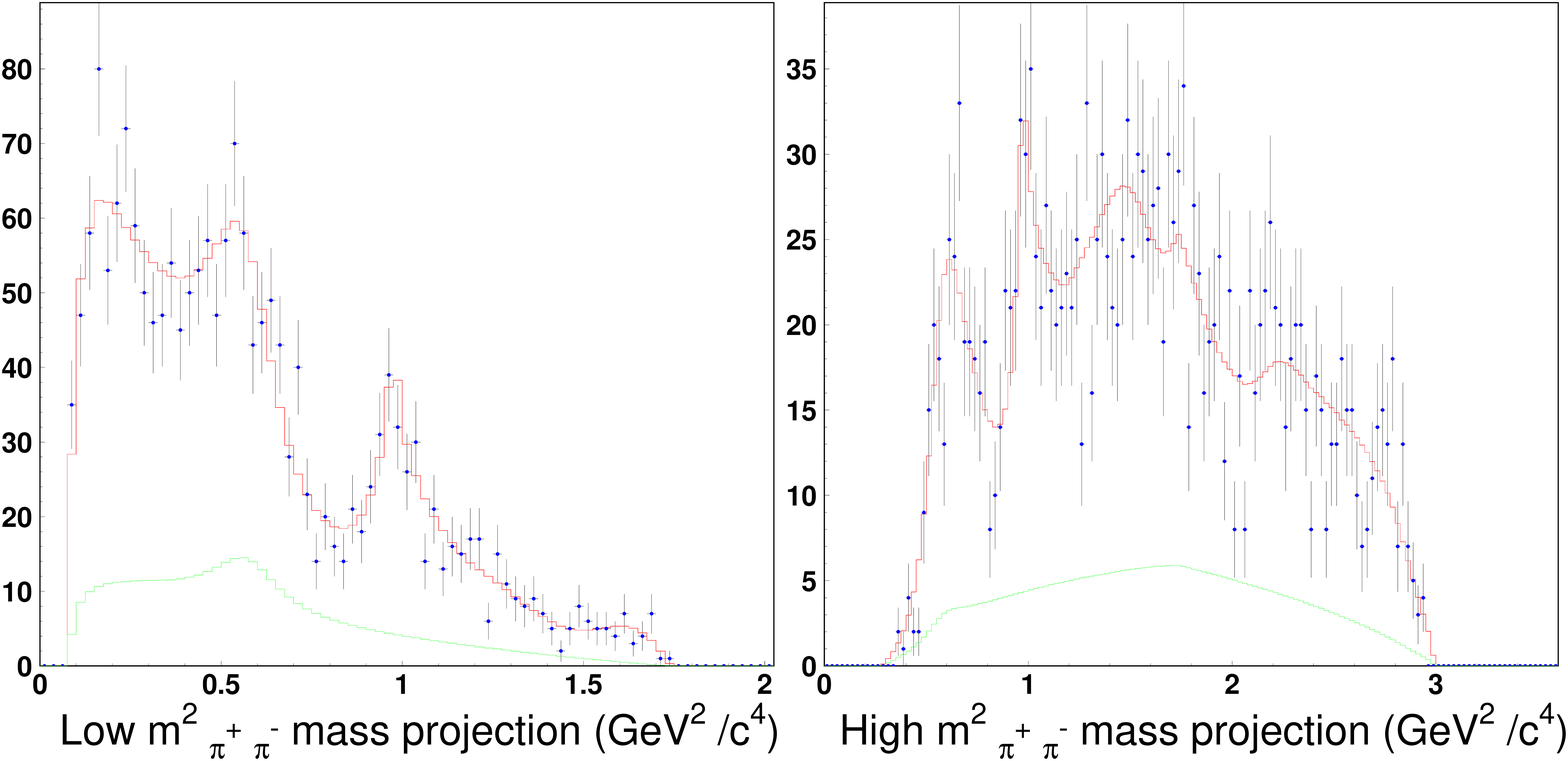} \caption{
      FOCUS Dalitz-plot projections for $D_s$ and $D^+$to three pions with fit results
      superimposed. The background shape under the signal is also shown.}
    \label{d_proj_Kmatrix} 
\end{figure}

The FOCUS collaboration has implemented the \emph{K-matrix} approach in the
$D_s$ and $D^+ \to \pi^+\pi^-\pi^+$ analyses. Results and details can be found
in \cite{Focus_kmat}. It was the first application of this formalism in the
charm sector. In this model \cite{aitch}, the production process, i.e, the D
decay, can be viewed as consisting of an initial preparation of states,
described by the \emph{P-vector}, which then propagates according to
\mbox{$(I-iK\rho)^{-1}$} into the final one. The \emph{K-matrix} here is the
scattering matrix and is used as fixed input in our analysis. Its form was
inferred by the global fit to a rich set of data performed in \cite{anisar1}.
It is interesting to note that this formalism, beside restoring the proper
dynamical features of the resonances, allows for the inclusion in $D$ decays
of the knowledge coming from scattering experiments, i.e, an enormous amount
of results and science. No re-tuning of the \emph{K-matrix} parameters was
needed. The confidence levels of the final fits are 3.0\,\%  and 7.7\,\% for
the $D_s$ and $D^+$ respectively. The results were extremely encouraging since
the same \emph{K-matrix} description gave a coherent picture of both two-body
scattering measurements in light-quark experiments \emph{as well as}
charm-meson decay. This result was not obvious beforehand. Furthermore, the
same model was able to reproduce features of the $D^+\to\pi^+\pi^-\pi^+$
Dalitz plot, shown in  fig.\ref{d_proj_Kmatrix}, that would otherwise require
an \emph{ad hoc} $\sigma$ resonance. The better treatment of the $S$-wave
contribution provided by the \emph{K-matrix} model was able reproduce the
low-mass $\pi^+\pi^-$ structure of the $D^+$ Dalitz plot. This suggests that
any $\sigma$-like object in the $D$ decay should be consistent with the same
$\sigma$-like object measured in $\pi^+\pi^-$ scattering.

Further considerations and conclusions from the FOCUS three-pion analysis were
limited by the sample statistics, i.e. $1475 \pm 50$ and $1527 \pm 51$ events
for $D_s$ and $D^+$ respectively.

We considered imperative to test the formalism at higher statistics. This was
accomplished by the $D^+ \to K^-\pi^+\pi^+$ analysis. 

\section{ The $D^+ \to K^- \pi^+ \pi^+$ amplitude analysis} 

\begin{figure}[!ht] \centering
  {
 \includegraphics[width=65mm]{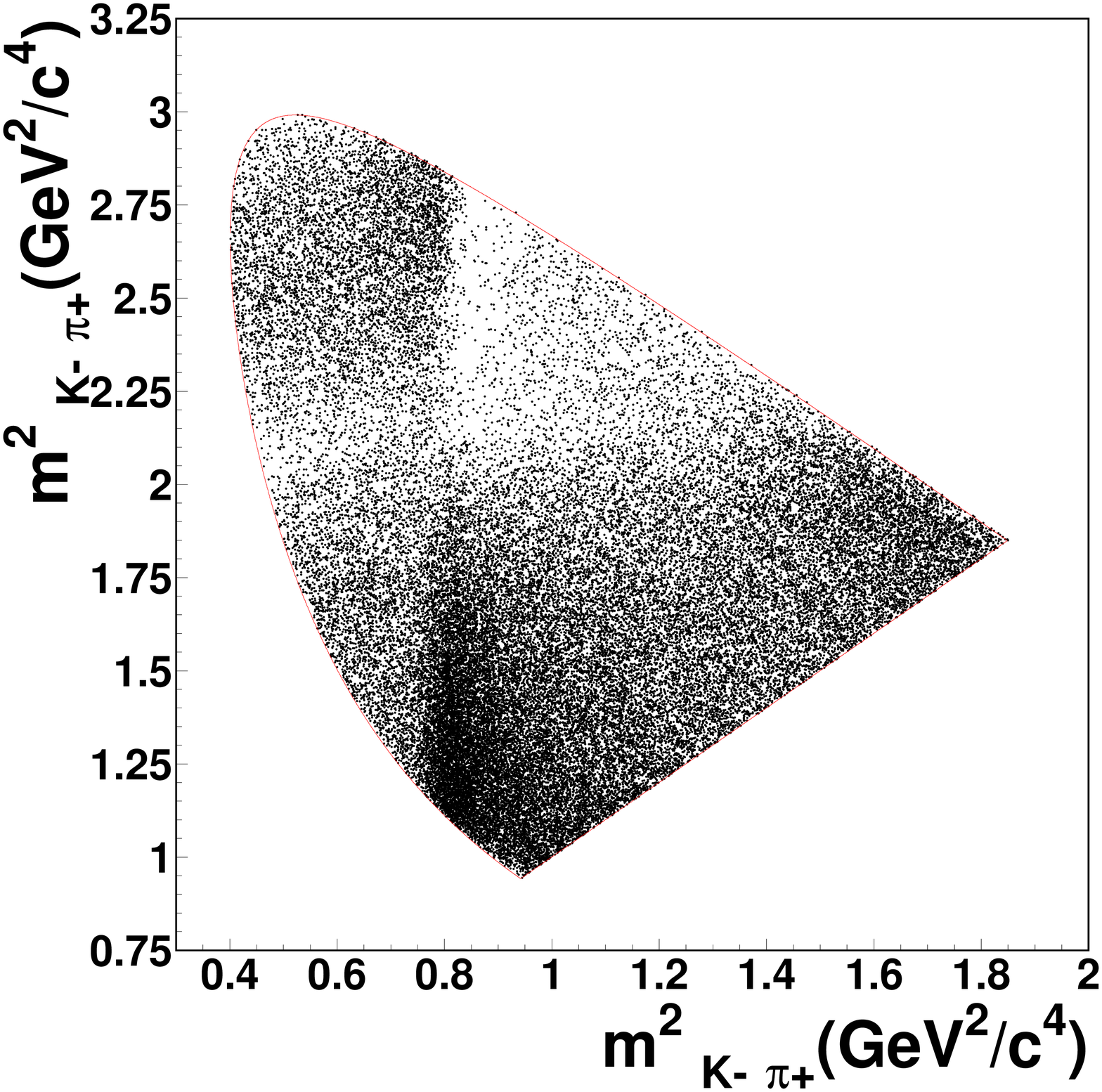}
 \includegraphics[width=67mm]{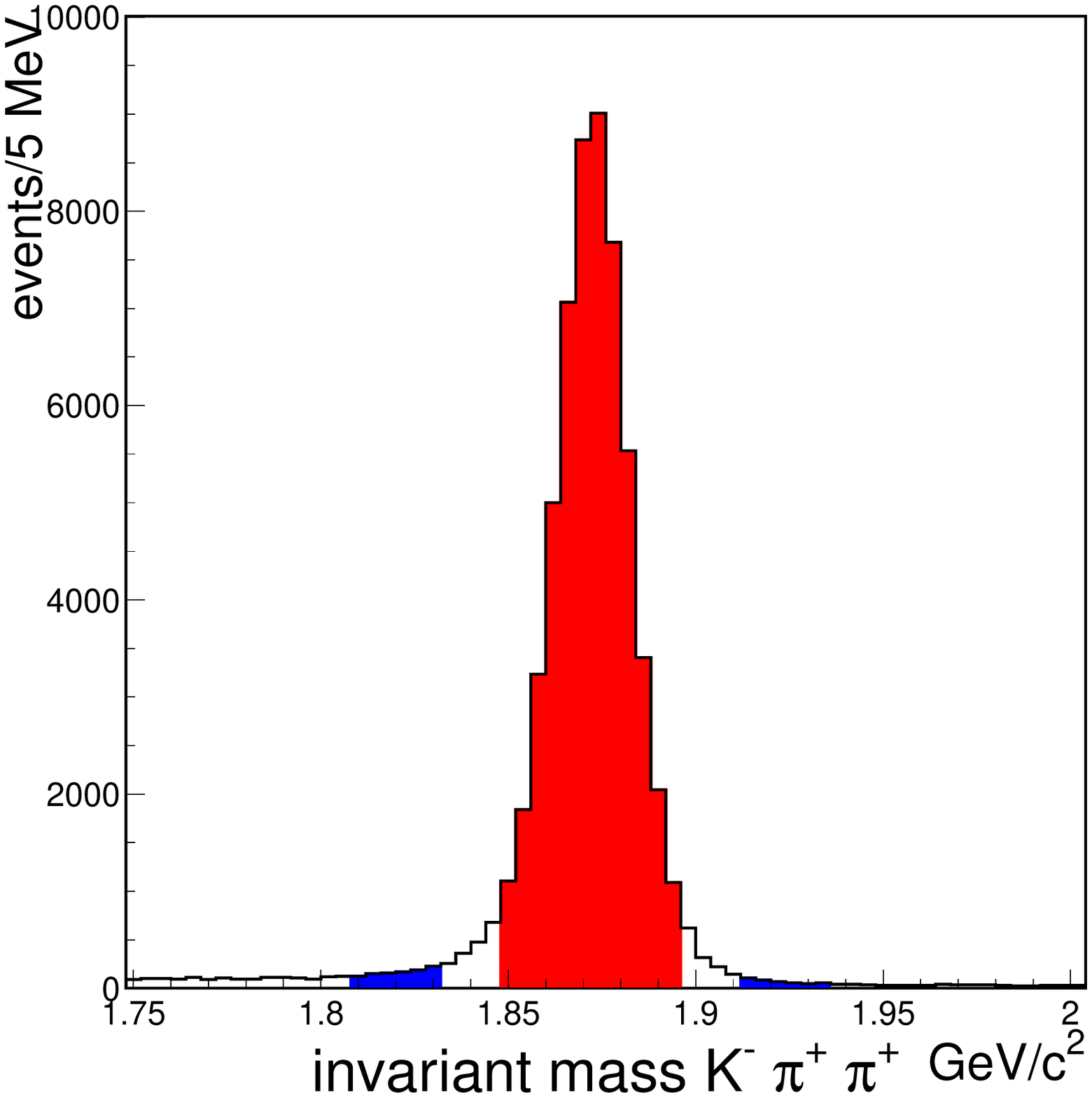}
}
 \caption{The $D^+ \to K^-\pi^+\pi^+$ Dalitz plot (left) and mass distribution (right). Signal and sideband regions
 are indicated: sidebands are at $\pm$(6--8)
 $\sigma$ from the peak.}
\label{signal}
\end{figure}

The recent FOCUS  study
of the $D^+ \to K^-\pi^+\pi^+$ channel uses 53653 Dalitz-plot events with a
signal fraction of $\sim$ 97\%, and represents the 
highest statistics, most
complete Dalitz plot published analysis for this channel. Invariant mass and Dalitz
plots are shown in fig.\ref{signal}. Details of the analysis may be found in
\cite{Focus_kpp}.

An additional complication in the $K\pi$ system comes from the presence in the
$S$-wave of the two isospin states, $I=1/2$ and $I=3/2$. Although only the
$I=1/2$ is dominated by resonances, both isospin components are involved in
the decay of the $D^+$ meson into $K^-\pi^+\pi^+$. A model for the decay
amplitudes of the two isospin states can be constructed from the 2 $\times$ 2
\emph{K-matrix} describing the $I=1/2$ $S$-wave scattering in $(K\pi)_1$ and
$(K\eta')_2$ (with the subscripts 1 and 2, respectively, labelling these two
channels), and the single-channel \emph{K-matrix} describing the $I=3/2$
$K^-\pi^+ \to K^-\pi^+$ scattering. The \emph{K-matrix} form we use as input
describes the $S$-wave $K^-\pi^+ \to K^-\pi^+$ scattering from the LASS
experiment \cite{lass} for energy above 825 MeV and $K^-\pi^- \to K^- \pi^-$
scattering from Estabrooks \emph{et al.} \cite{estabrook}. The \emph{K-matrix}
form follows the extrapolation down to the $K\pi$ threshold for both $I=1/2$
and $I=3/2$ $S$-wave components by the dispersive analysis by B\"uttiker
\emph{et al.} \cite{butt}, consistent with Chiral Perturbation Theory
\cite{cpt}. The total $D$-decay amplitude can be written as

\begin{equation}
\mathcal{M}= {(F_{1/2})}_1(s) + F_{3/2}(s) + \sum_j a_j~
e^{i\delta_j}~B(abc|r), \label{A_tot_kmat}
\end{equation}
\par\noindent
where $s=M^2(K\pi)$, ${(F_{1/2})}_1$ and $F_{3/2}$ represent the $I=1/2$ and
$I=3/2$ decay amplitudes in the $K\pi$ channel, $j$ runs over vector and
spin-2 tensor resonances and $~B(abc|r)$ are Breit--Wigner forms. The $J>0$
resonances should, in principle, be treated in the same \emph{K-matrix}
formalism. However, the contribution from the vector wave comes mainly from
the $K^*(892)$ state, which is  well separated from the higher mass
$K^*(1410)$ and $K^*(1680)$, and the contribution from the spin-2 wave comes
from $K_2^*(1430)$ alone. Their contributions are limited to small
percentages, and, as a first approximation, they can be reasonably described
by a simple sum of Breit--Wigners. More precise results would require a better
treatment of the overlapping $K^*(1410)$ and $K^*(1680)$ resonances as well.
In accordance with SU(3) expectations, the coupling of the $K\pi$ system to
$K\eta$ is supposed to be suppressed. Indeed we find little evidence that it
is required. Thus the $F_{1/2}$ form for the $K\pi$ channel is

\begin{equation}
(F_{1/2})_1= (I-iK_{1/2}\rho)_{1j}^{-1}(P_{1/2})_j,
 \label{F_12}
\end{equation}
\par\noindent
where $I$ is the identity matrix, $K_{1/2}$ is the \emph{K-matrix} for the
$I=1/2$ $S$-wave scattering in $K\pi$ and $K\eta'$, $\rho$ is the
corresponding phase-space matrix for the two channels \cite{chung} and
$(P_{1/2})_j$ is the production vector in the channel $j$.

The form for $F_{3/2}$ is
\begin{equation}
F_{3/2}= (I-iK_{3/2}\rho)^{-1}P_{3/2},
 \label{F_32}
\end{equation}
\par\noindent
where $K_{3/2}$ is the single-channel scalar function describing the $I=3/2$
\, \mbox{$K^-\pi^+ \to K^-\pi^+$} scattering, and $P_{3/2}$ is the production
function into $K\pi$.

The \emph{P-vectors} are in general complex reflecting the fact that the
initial coupling $D^+\to (K^- \pi^+)\pi^+_{spectator}$ need not be real. Their
functional forms are:

\begin{equation}
(P_{1/2})_1= \frac{\beta g_1 e^{i\theta}} {s_1-s}
 + ( c_{10} + c_{11}\widehat{s}  + c_{12}
{\widehat{s}}^2 ) e^{i\gamma_1} \label{P_121}
\end{equation}

\begin{equation}
(P_{1/2})_2= \frac{\beta g_2 e^{i\theta}} {s_1-s} + (c_{20} + c_{21}\widehat
{s} +c_{22}{\widehat{s}}^2)e^{i\gamma_2} \label{P_122}
\end{equation}

\begin{equation}
P_{3/2}= (c_{30} + c_{31}\widehat{s} +c_{32}{\widehat{s}}^2)e^{i\gamma_3}
 \label{P_32}.
\end{equation}

\par\noindent
$\beta e^{i\theta} $ is the complex coupling to the pole in the `initial'
production process, $g_1$ and $g_2$, $s_1$ and $s_2$  are the \emph{K-matrix}
couplings and poles. The polynomials are expanded about $\widehat s = s -s_c$,
with \mbox{$s_c = 2$ \, GeV$^2$} corresponding to the center of the Dalitz
plot. The polynomial terms in each channel are chosen to have a common phase
$\gamma_i$ to limit the number of free parameters in the fit and avoid
uncontrolled interference among the physical background terms. Coefficients
and phases of the \emph{P-vectors} are the only free parameters of the fit
determining the scalar components. $K\pi$ scattering determines the parameters
of the {\it K-matrix} elements and these are fixed inputs to this $D$-decay
analysis. Free parameters for vectors and tensors are amplitudes and phases
($a_i$ and $\delta_i$). Table ~\ref{tab_P_vector_nov06} reports our
\emph{K-matrix} fit results. It shows that quadratic terms in $(P_{1/2})_1$
are significant in fitting data, while in both  $(P_{1/2})_2$ and $P_{3/2}$
constants are sufficient. The $J>0$ states required by the fit are listed in
table~\ref{tab_high_spin}.
\begin{table}[!htb]
 \centering
  \caption{$S$-wave  parameters from the \emph{K-matrix} fit
 to the FOCUS $D^+ \to K^-\pi^+\pi^+$ data.  The first error is statistic, the second error is systematic from the experiment,
 and the third is systematic induced by model input parameters for higher resonances.
 Coefficients are for the unnormalized $S$-wave.}
 \label{tab_P_vector_nov06}
 \begin{tabular}{cc}
 \hline
 \bf{coefficient} & \bf{phase (deg})\\
 \hline
  $\beta$      = $3.389  \pm  0.152 \pm 0.002  \pm 0.068   $     & $\theta=    286 \pm 4 \pm 0.3 \pm 3.0   $  \\
 $c_{10}$     = $1.655  \pm  0.156 \pm 0.010  \pm 0.101   $     & $\gamma_1 = 304 \pm 6 \pm 0.4 \pm 5.8$ \\
 $c_{11}$     = $0.780  \pm  0.096 \pm 0.003  \pm 0.090   $     &      \\
 $c_{12}$     = $-0.954 \pm 0.058  \pm  0.0015 \pm 0.025  $     &      \\
 $c_{20}$     = $17.182 \pm 1.036  \pm 0.023 \pm 0.362    $     & $ \gamma_2 = 126  \pm  3 \pm 0.1 \pm 1.2  $ \\
 $c_{30}$     = $0.734  \pm 0.080  \pm 0.005 \pm 0.030    $     & $ \gamma_3 = 211  \pm 10 \pm 0.7\pm 7.8 $ \\
 \hline
 \hline
 \multicolumn{2}{c}{
{Total $S$-wave fit fraction}  = $ 83.23 \pm 1.50 \pm 0.04 \pm 0.07$ \%}  \\
 \multicolumn{2}{c}{
{Isospin 1/2 fraction }  = $207.25 \pm 25.45 \pm  1.81 \pm 12.23$ \% } \\
 \multicolumn{2}{c}{
{Isospin 3/2 fraction }  = $40.50  \pm 9.63 \pm 0.55 \pm 3.15$ \% } \\
 \hline
 \end{tabular}
\end{table}
\begin{table}[!htb]
\centering
 \caption{Fit fractions, phases, and coefficients for the $J>0$ components from the \emph{K-matrix} fit
 to the FOCUS  $D^+ \to K^-\pi^+\pi^+$ data.  The first error is statistic, the
 second error is systematic from the experiment, and the
 third error is systematic induced by model input parameters for higher resonances.
 }
 \label{tab_high_spin}
 \begin{tabular}{cccc}
 \hline
 \bf{component} & \bf{fit fraction (\%)} & \bf{phase $\delta_j$ (deg)}
 & \bf{coefficient} \\
 \hline
 $K^*(892)\pi^+$     &  $13.61 \pm  0.98 $       &    0 (fixed)                          & 1 (fixed)          \\
                     &  $ \pm \ 0.01 \pm 0.30 $  &                                       &                    \\

 $K^*(1680)\pi^+$    &  $ 1.90 \pm  0.63  $      &  $1   \pm  7  $       & $0.373 \pm 0.067 $ \\
                     &  $ \pm \ 0.009 \pm 0.43$  &  $ \pm \ 0.1 \pm 6 $  & $ \pm \ 0.009 \pm 0.047$ \\
 $K^*_2(1430)\pi^+$  &  $ 0.39 \pm  0.09  $      &  $296 \pm 7   $       & $0.169 \pm 0.017$ \\
                     &  $\pm \ 0.004 \pm 0.05 $  & $\pm\  0.3 \pm 1 $    & $ \pm \ 0.010 \pm 0.012 $ \\
 $K^*(1410)\pi^+$    &  $ 0.48 \pm  0.21  $      &  $293 \pm 17  $       & $0.188 \pm 0.041 $ \\
                     &  $ \pm \ 0.012 \pm 0.17$  & $\pm \ 0.4 \pm 7 $    & $ \pm \ 0.002 \pm 0.030$ \\
 \hline
 \end{tabular}
\end{table}

\begin{figure}[h]
 \centering
 \includegraphics[width=0.8\textwidth]{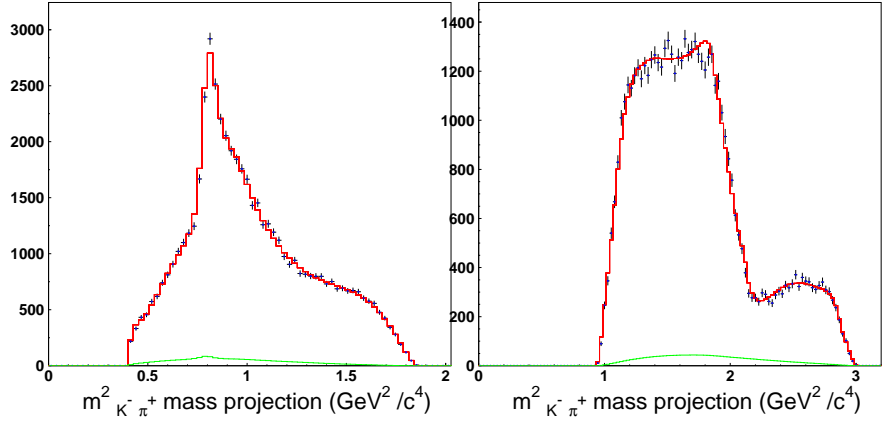}
 \caption {The Dalitz plot projections with the \emph{K-matrix} fit superimposed.
  The background shape under the signal is also shown.}
 \label{fit_kmatrix_proj}
\end{figure}

The $S$-wave component accounts for the dominant portion of the decay $(83.23
\pm 1.50) \%$. A significant fraction, $13.61\pm 0.98$\%, comes, as expected,
from $K^*(892)$; smaller contributions come from two vectors $K^*(1410)$ and
$K^*(1680)$ and from the tensor $K_2^*(1430)$. It is conventional to quote fit
fractions for each component and this is what we do. Care should be taken in
interpreting some of these since strong interference can occur. This is
particularly apparent between contributions in the same-spin partial wave.
While the total $S$-wave fraction is a sensitive measure of its contribution
to the Dalitz plot, the separate fit fractions for $I=1/2$ and $I=3/2$ must be
treated with care. The broad $I=1/2$ $S$-wave component inevitably interferes
strongly with the slowly varying $I=3/2$ $S$-wave, as seen for instance in
\cite{mike_laura}. Fit results on the projections are shown in
fig.~\ref{fit_kmatrix_proj}.

\begin{figure}[h]
 \centering
\includegraphics[width=80mm]{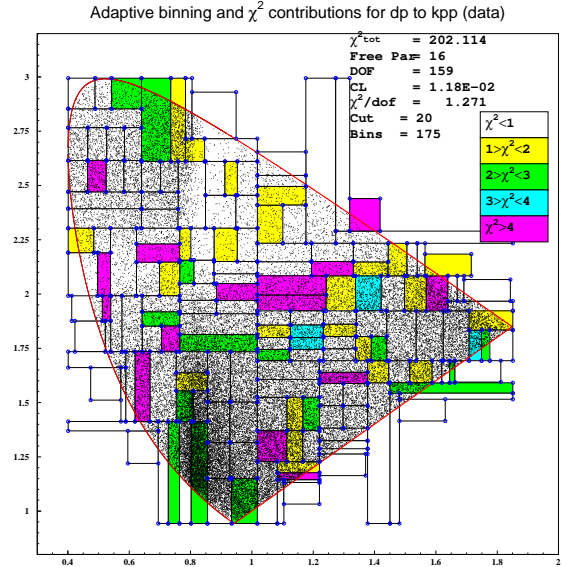}
\caption{ The adaptive binning scheme. }
  \label{fit_adapt}
\end{figure}

The fit $\chi^2$/d.o.f is 1.27 corresponding to a 
confidence level of 1.2\%. Our adaptive binning scheme is shown is fig~\ref{fit_adapt} 
If the $I=3/2$ component is removed from the fit, the $\chi^2$/d.o.f worsens
to 1.54, corresponding to a confidence level of $10^{-5}$.

\begin{figure}[h]
\centering
  \includegraphics[width=0.8\textwidth]{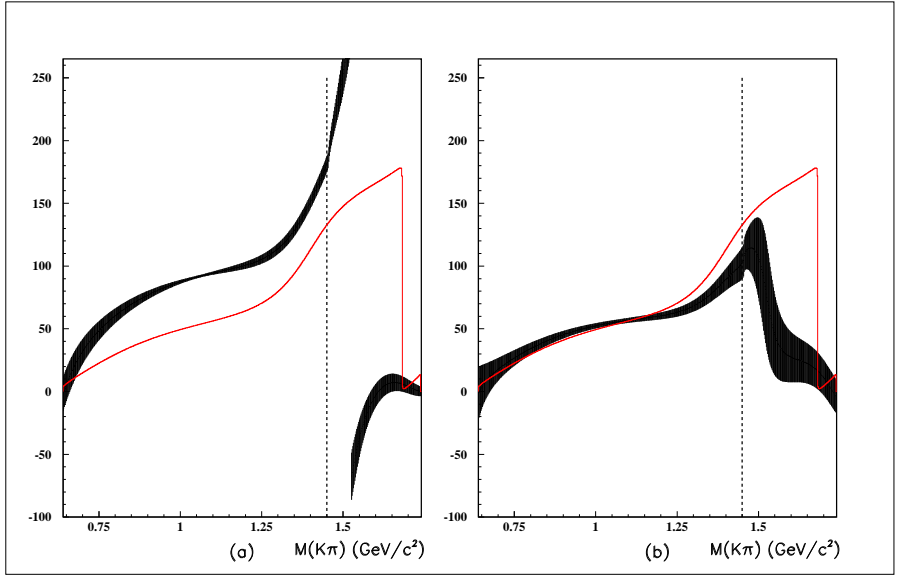}
  \caption{Comparison between the LASS $I=1/2$ phase + ChPT (continous line)
  and the \emph{F-vector} phases (with $\pm 1\,\sigma$ statical error bars); a)
   total \emph{F-vector} phase; b) $I=1/2$ \emph{F-vector} phase.
  The vertical dashed line shows the location of the $K\eta'$.}
  \label{phase_tot_compa}
\end{figure}

These results can be compared with those obtained in the effective isobar
model, consisting in a sum of Breit Wigners, which can serve as the standard
for fit quality.  Two \emph{ad hoc} scalar resonances are required, of mass
$856 \pm 17$ and $1461 \pm 4$ and width $464 \pm 28$ and $177 \pm 8$ MeV/$c^2$
respectively to reproduce the data and reach a $\chi^2$/d.o.f is 1.17,
corresponding to a C.L of 6.8\%. A detailed discussion of the results and the
systematics can be found in \cite{Focus_kpp}. A feature of the \emph{K-matrix}
amplitude analysis is that it allows an indirect phase measurement of the
separate isospin components: it is this phase variation with isospin $I=1/2$
that should be compared with the same $I=1/2$ LASS phase, extrapolated from
825 GeV down to threshold according to Chiral Perturbation Theory. This is
done in the right plot of fig.~\ref{phase_tot_compa}. In this model
\cite{aitch} the \emph{P-vector} allows for a phase variation accounting for
the interaction with the third particle in the process of resonance formation.
It so happens that the Dalitz fit gives a nearly constant production phase.
The two phases in fig.~\ref{phase_tot_compa}b) have the same behaviour up to
$\sim$ 1.1 GeV. However, approaching $K\eta'$ threshold, effects of
inelasticity and differing final state interactions start to appear. The
difference between the phases in fig.~\ref{phase_tot_compa}a) is due to the
$I=3/2$ component.

These results are consistent with $K\pi$ scattering data, and consequently
with Watson's theorem predictions for two-body $K\pi$ interactions in the low
$K\pi$ mass region, up to $\sim$ 1.1 GeV, where elastic processes dominate.
This means that possible three-body interaction effects, not accounted for in
the \emph{K-matrix} parametrization, play a marginal role. The \emph{K-matrix}
form used in this analysis generates the \emph{S-matrix} pole $E=M
-i\Gamma/2=1.408 -{\it i} 0.110$ GeV. Any more distant pole than $K_0^*(1430)$
is not reliably determined as this simple \emph{K-matrix} expression does not
have the required analyticity properties. However, our {\it K-matrix}
representation fits along the real energy axis inputs on scattering data and
Chiral Perturbation Theory in close agreement with those used in
\cite{Desco_Mussalla}, which locates the $\kappa$ with a mass of $(658 \pm
13)$ MeV and a width of $(557 \pm 24)$ MeV by careful continuation. These pole
parameters  are quite different from those implied by the simple isobar fits.
We have thus shown that whatever $\kappa$ is revealed by our $D^+\to
K^-\pi^+\pi^+$ results, it is the same as that found in scattering data.

\section{A non-parametric approach to determine the $K^-\pi^+$ amplitude in
$D^+\to K^-K^+\pi^+$ decay}

While making the effort of refining the amplitude formalism, FOCUS identified
the $D^+ \to K^-K^+\pi^+$ as an ideal case to apply the projective weighting
technique developed in the semi-leptonic sector \cite{Focus_projective_sl} to
the hadronic decays, with no need to assume specific Breit-Wigner resonances,
forms for mass dependent widths, hadronic form factors or Zemach momentum
factors. 
Details can be found in \cite{Focus_projective_had}. The
old E687 Dalitz plot analysis \cite{e687_kkpi} concluded that the observed
$D^+ \to K^-K^+ \pi^+$ Dalitz plot could be adequately described by just three
resonant contributions: $\phi \pi^+, K^+ \bar K^*(892)$ and
$K^+\bar{K^*}_0(1430)$. Although $\phi \pi^+$ is an important contribution,
the $\phi$ is a very narrow resonance that can be substantially removed
through a cut on $m_{K^+K^-}$, i.e $m_{K^+K^-} >$ 1050 MeV/c$^2$. Since there
is no overlap of the $\phi$ band with the $\bar K^*$ and most of the
kinematically allowed $\bar {K^*}_0(1430)$ region, there is a relatively small
loss of information from the anti-$\phi$ cut; of course careful systematic
evaluation for residual $K^+K^-$ contributions and bias are performed. In the
absence of the $K^-K^+$ resonances, we can write the decay amplitude in terms
of $m_{K^-\pi^+}=m$ and the helicity decay angle $\theta$ Thus
\begin{equation}
A=\sum_{l}^{s,p.d...} A_l(m)d^l_{00}(\cos\theta), \label{non_para}
\end{equation}
where $d^l_{00}(cos\theta)$ are the Wigner d-matrices describing the amplitude
for a $K^-\pi^+$ system of angular momentum $l$ to simultaneously have 0
angular momentum along its helicity axis and the $K^-\pi^+$ decay axis. This
technique is an intrinsically one-dimensional analysis.  The decay intensity
assuming, for simplicity, that only S and P-waves are present, is

\begin{equation}
|A|^2 = |S(m) +P(m)\cos\theta|^2 =|S(m)|^2 +2
\mathrm{Re}\{S^*(m)P(m)\}\cos\theta +|P(m)|^2 \cos^2\theta ,
\label{non_para_inte}
\end{equation}

The approach is to divide $\cos\theta$ into twenty evenly spaced angular bins.
Let
\begin{equation}
\overrightarrow{D} = (^in_1,^in_2...^in_{20})
 \label{dvector}
\end{equation}
\noindent
 be a vector whose 20 components give the population in data for
each of the 20 $\cos\theta$ bins. Here $^i$ specifies the $i^{th}
m_{K^-\pi^+}$ bin. Our goal is to represent the $\overrightarrow{D}$ vector in
eq.~\ref{dvector} as a sum over the expected populations for each of the three
partial waves. For this simplified case there are three such vectors computed
for each $m_{K^-\pi^+}$ bin, $\{^i
 \overrightarrow {m}_{\alpha} \} = (^i \overrightarrow{m}_{SS}, ^i \overrightarrow{m}_{SP}, ^i \overrightarrow{m}_{PP}
 )$.
  \begin{figure}[h]
 \centering
 \includegraphics[width=0.8\textwidth]{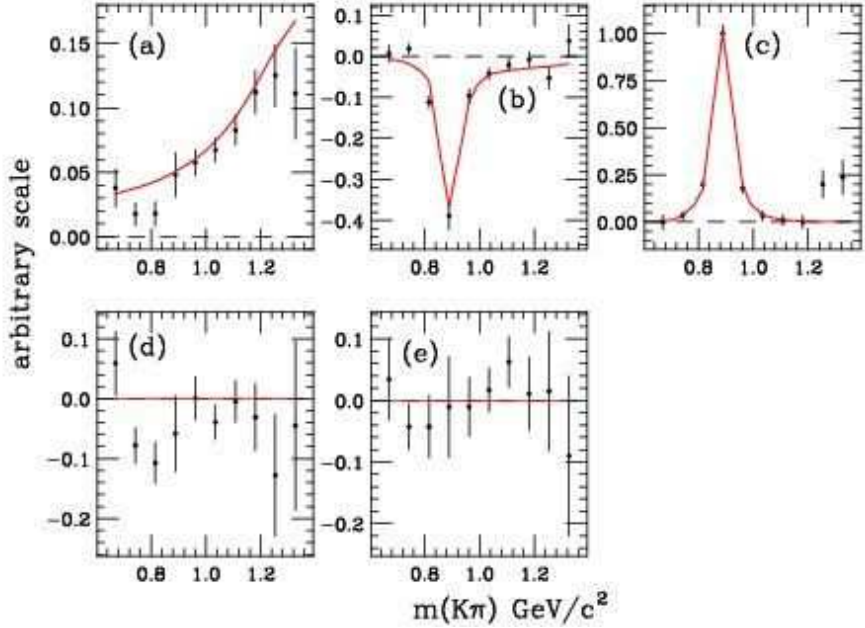}
 \caption{This figure compares the five projected amplitudes obtained according to their angular dependence (error bars) with the E687 model properly
 corrected, as explained in the text, to match the data (red curves). The plots are: a) S$^2$ direct term, b) 2 S$\times$P interference term,
c) P$^2$ direct term, d)2 P$\times$D interference term and e) D$^2$ direct
term.}
 \label{projector}
 \end{figure}

 Each $^i\overrightarrow {m}_{\alpha}  $ is generated using a phase-space and
 full detector simulation for $D^+ \to K^-K^+\pi^+$ decay with one amplitude
 turned on, and all the others turned off. We then use a weighting technique to
 fit the bin populations in the data to the form $ ^i\overrightarrow{D} =
 F_{SS}(m_i) ^i \overrightarrow{m}_{SS} + F_{SP}(m_i) ^i
 \overrightarrow{m}_{SP}+ F_{PP}(m_i)^i\overrightarrow{m}_{PP}$. When
 including the D-wave as well, the results appear just as five weighted histograms in the
 $m_{K^-\pi^+}$ mass, as in fig.~\ref{projector}, for the five independent amplitude
 contributions.

 The curves in fig.~\ref{projector} are the model used in E687 but with a $\bar {K^*}_0(1430)$
 \emph{ad hoc} arranged to fit the data, i.e represented as a Breit Wigner  with a pole at $m_0$=1412 MeV/$c^2$ and a
 width of $\Gamma$ = 500 MeV/$c^2$, not consistent with the standard PDG
 $\bar {K^*}_0(1430)$ parametrization used by E687. This analysis reveals once more, how subtle the inclusion of the broad
 S-wave resonances in charm Dalitz analysis can be.
 Although the $D^+ \to K^-K^+\pi^+$ is an ideal case, it might be possible to
 extend the analysis to the $D_s \to K^-K^+\pi^+$ decay, as well as $D^0 \to
 K^+K^-\bar K^0$ and hadronic four body decays such as $D^0 \to K^-K^+
 \pi^+\pi^- \to \phi \pi^+\pi^-$.

\section{Search for pentaquark candidates}

The FOCUS collaboration searched for the charmed  $\Theta_c^0(\bar c uudd)$
pentaquark candidate in the decay modes  $\Theta_c^0 \to D^{*-}p$ and
$\Theta_c^0 \to D^-p$ \cite{Focus_penta_1}. No evidence for a pentaquark at
3.1 GeV/$c^2$ or at any mass less than 4 GeV/$c^2$ was observed. More recently
the search was extended to two other candidates: $\Theta^+(\bar suudd) \to
pK^0_s $ \cite{Focus_penta_2} and $\phi^{--}(1860)(ssddu) \to \Xi^-\pi^-$
\cite{Focus_penta_3}. Having found no evidence, limits were calculated. The
$\Theta^+$ production cross section was normalized to $\Sigma^*(1385)^{\pm}$
and $K^*(892)^+$ because the reconstructed decay modes of the particles
$\Sigma^*(1385)^{\pm} \to \Lambda^0 \pi^{\pm}$ and $K^*(892)^+ \to K_S^0
\pi^+$ are very similar, in terms of topology and energy release, to the
signal. The 95 \% C.L upper limits of $\frac{\sigma(\Theta^+) \cdot
BR(\Theta^+ \to pK_S^0)}{\sigma(K^*(892)^+)} <$ 0.00012 (0.00029) and $
\frac{\sigma(\Theta^+)\cdot BR(\Theta^+ \to
pK_S^0)}{\sigma(\Sigma^*(1385)^{\pm})} < $ 0.0042 (0.0099) were  estimated for
a natural width of 0 and 15 MeV/$c^2$ in the good acceptance region of the
detector , i.e. for parent particles with momenta above 25 GeV/c. Analogously
the upper limit was calculated for the $\Xi^{--}_5 (\phi^{--}(1860))$
candidate with respect to the $\Xi^*(1530)^0 \to \Xi^- \pi^+$ obtaining
$\frac{\sigma(\Xi^{--}_5)\cdot BR(\Xi^{--}_5 \to \Xi^-
\pi^-)}{\sigma(\Xi^*(1530)^0} <$ 0.007 (0.019) for a natural width of 0 (15)
MeV/$c^2$.

\section{Conclusions}

Dalitz-plot analysis represents a unique, powerful and promising tool for
physics studies within and beyond the Standard Model; however to perform such
sophisticate analyses, we need to model the strong interaction effects. FOCUS
 has performed pilot studies in the charm sector through the
\emph{K-matrix} formalism and has started an effort to identify channels where
non-parametric approaches can be undertaken. What has been learnt from charm
will be beneficial for future accurate beauty measurements.
%


\end{document}